\documentstyle[11pt,newpasp,twoside,psfig]{article}
\index{pulsar wind nebulae!hydrodynamics}
\index{pulsar wind nebulae!interactions with supernova remnants}

\def\edcomment#1{\iffalse\marginpar{\raggedright\sl#1\/}\else\relax\fi}
\reversemarginpar

\begin{document}
\title{A Key to Pulsar Wind Bubble Morphologies: HD simulations}
 \author{Eric van der Swaluw}
\affil{FOM-Institute for Plasma Physics Rijnhuizen, P.O. Box 1207, 3430 BE Nieuwegein, The Netherlands}
\author{Turlough P. Downes}
\affil{School of Mathematical Sciences, Dublin City University, Glasnevin, Dublin 9, Ireland}

\begin{abstract}

We present a model of a pulsar-driven supernova remnant, by using a 
hydrodynamics code, which simulates the evolution of a pulsar wind nebula 
when the pulsar is moving at a high velocity through its expanding supernova 
remnant. The simulation shows four different stages of
the pulsar wind nebula: the supersonic expansion stage, the reverse shock
interaction stage, the subsonic expansion stage and ultimately the bow shock
stage. Due to the high velocity of the pulsar, the position of the pulsar is 
located at the head of the pulsar wind bubble, after the passage of the 
reverse shock. The resulting morphology of the pulsar wind bubble is therefore
similar to the morphology of a bow shock pulsar wind nebula. We show how to
distinguish these two different stages, and apply this method to the SNR 
G327.1-1.1, for which we argue that there is {\it no} bow shock around its
pulsar wind nebula.
\end{abstract}

\section{Introduction}

The dynamics of the interior of a young pulsar-driven supernova remnant (SNR)
is dominated by the continuous injection of energetic particles by a 
relativistic pulsar wind. This pulsar wind is driven by the spin-down energy 
of the pulsar, and is terminated by a strong MHD shock (Rees \& Gunn 1974). 
The pulsar wind blows a pulsar wind nebula (PWN), which is bounded by a strong
PWN shock, into the freely expanding ejecta of its 
surrounding young SNR. A young SNR is characterised by a blastwave propagating 
into the interstellar medium (ISM) and a reverse shock, which propagates back 
into the SNR interior once the SNR blastwave has swept up a few times the 
ejecta mass (McKee \& Truelove 1995). When the reverse shock collides with 
the PWN shock, the supersonic expansion stage of the PWN is terminated: the 
PWN shock bounding the hot pulsar wind bubble disappears. 

Hydrodynamical simulations of the above process have been performed by
several authors (van der Swaluw et al. 2001, Blondin et al. 2001) for 
a centered pulsar. These simulations bear out that the timescale for the
reverse shock interaction stage is comparable with the lifetime of the 
supersonic expansion stage. Ultimately the expansion of the PWN proceeds 
subsonically inside the relaxed Sedov-Taylor SNR, when the reverberations 
of the reverse shock have vanished.

In this paper we present a hydrodynamical simulation of a PWN when the 
pulsar is {\it moving} at a high velocity through the expanding SNR. The 
simulation shows that due to the high velocity of the pulsar, the position 
of the pulsar is off-centered with respect to its PWN, after the passage of 
the reverse shock.
Furthermore, the simulation shows a deformation of the PWN 
into a bowshock when the motion of the pulsar becomes supersonic.  This 
occurs at half the crossing time, or equivalently when $R_{\rm psr}/R_{\rm 
snr} \approx 0.677$ where $R_{\rm psr}$ is the distance of the pulsar from 
the center of the SNR, and $R_{\rm snr}$ is the radius of the 
blastwave. The crossing time indicates 
the age of the SNR when the pulsar overtakes the shell of its remnant, 
while the latter is in the Sedov-Taylor stage. Both values are in
complete agreement with analytical work performed by van der Swaluw 
et al. (1998).

\section{The evolution of a PWN inside a SNR}

\subsection{The Hydrodynamical Simulation}

We use a second order, properly upwinded hydrodynamics code (described
in Downes \& Ray 1999) to simulate the dynamics of the interaction 
between a pulsar wind and a supernova remnant.  The hydrodynamics equations 
are intergrated in cylindrical symmetry, and the boundary conditions are 
taken as gradient zero everywhere except on the $r=0$ boundary, where they 
are set to reflecting. The simulation is performed in the rest frame of the 
pulsar, which is moving at a velocity of $V_{\rm psr}=1,000$ km/sec. 
Thermal energy is deposited at a constant rate in a 
small sphere centered around the pulsar, such that the pulsar wind luminosity
equals $L_{\rm pw}=10^{38}$ erg/s. Mass is also deposited into this region 
at a rate chosen such that the pulsar wind terminal velocity equals 
$v_{\infty}$=30,000 km s$^{-1}$. The supernova itself is modeled by
initialising a sphere of radius 0.25 pc, or 21 grid cells, with a high 
thermal energy and density such that the total energy contained in the sphere 
is $E_0=10^{51}$ ergs, while the mass is $M_{\rm ej}=3 M_{\odot}$. This means 
that the total injected energy by the pulsar wind during its stay in the SNR 
interior equals $E_{\rm pw}={L_{\rm pw}t_{\rm cr}}\ll E_0$, here $t_{\rm cr}$ 
is the crossing time. The surrounding medium has a density of $\rho_0=10^{-24}$g/cm$^3$. The total grid size equals 7 by 14 parsec with an 
associated 600 by 1200 grid cells.

\subsection{An off-centered Pulsar inside its PWN}

Figure 1 shows a logarithmic gray-scale representation of the density 
distribution of the PWN/SNR system at two different
stages. The upper profile shows the system, shortly after the passage of
the reverse shock. The simulation bears out that due to the high velocity
of the pulsar, the position of the pulsar is off-centered with respect to 
its surrounding PWN, after the passage of the reverse shock. 
A typical timescale for the age of such a system was given by van der 
Swaluw (2003):
\begin{equation}
t_{\rm rev} \; =\;  1,045 E^{-1/2}_{51}
\left({M_{\rm ej}\over M_\odot}\right)^{5/6}n_0^{-1/3}\;\;{\rm years}\; ,
\end{equation}
here $E_{51}$ is the total mechanical energy of the SNR in units of 
$10^{51}$ erg, $M_{\rm ej}$ is the ejecta mass and $n_0(=\rho_0/2.34
\times 10^{-24}$g/cm$^3$) is the ambient hydrogen number density.

The lower profile shows the PWN/SNR system after the formation of the bow
shock. In this case the position of the pulsar is again located at the head
of the PWN, but the conditions $R_{\rm psr}/R_{\rm snr}\ge 0.677$ and
\begin{figure}
\begin{center}
\centerline{\psfig{file=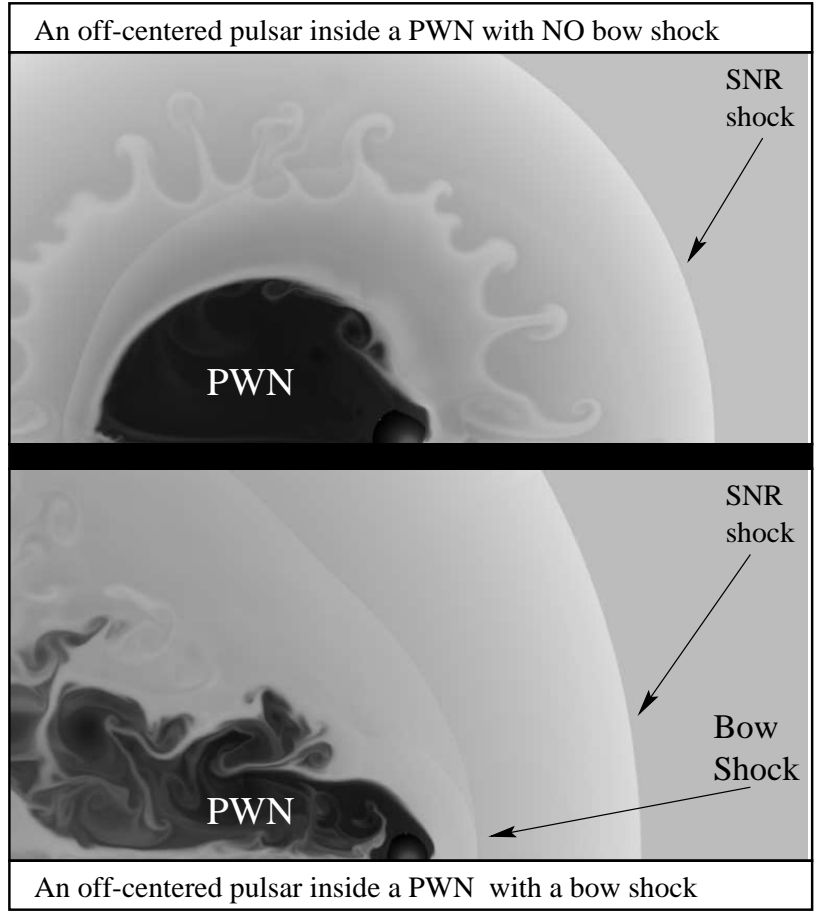,width=10.8 cm }}
\caption{The PWN/SNR system at an age of $t=2,600$ yrs (upper panel) and
$t=10,200$ yrs (bottom panel)}
\end{center}
\end{figure}
the age $t \ge 0.5 t_{\rm cr}$ are satisfied, where $t_{\rm cr}$ equals the
crossing time of the pulsar (van der Swaluw 2003):
\begin{equation}
t_{\rm cr}\;\simeq\; 1.4 \times 10^4 \; E_{51}^{1/3} 
V_{\rm 1000}^{-5/3}n_0^{-1/3} \;\; {\rm years} \; ,
\end{equation}
here $V_{1000}$ is the velocity of the pulsar in units of $1,000$ km/sec.

The most important issue of the present simulation for interpreting 
the evolutionary stage of plerionic components ( = the observational 
counterpart of a PWN) inside composite remnants, is the off-centered 
position of the pulsar inside its PWN. In order to determine whether
a PWN inside a composite remnant is in the {\it bow shock stage} one
should have $R_{\rm psr}/R_{\rm snr}\ge 0.677$ together with an age
which is larger than half the crossing time. If this is not the case
then the PWN is in either the {\it reverse shock interaction stage}
or the {\it subsonic expansion stage}. 

\subsection{Do we observe a bow shock in SNR G327.1-1.1 ... ?  }

Sun et al. (1999) presented a radio contour map of SNR G327.1-1.1, using 
MOST data overlied with X-ray data from ROSAT. The X-ray emission is centered 
around a finger of radio emission sticking out of a central radio bright 
region, indicating the presence of a pulsar wind. Following Sun et al. (1999) 
the SNR can be modelled in X-rays by the following set of parameters: 
$E_{51}=0.23$, $n_0=0.10$, $V_{\rm psr}$=600km/sec and an age of 
$t=1.1\times 10^4$. Using equation (2) to calculate $t_{\rm cr}$ we get a 
value of $4.3\times 10^4$ years: the age of the remnant is lower than 
half the crossing time. Furthermore the position of the PWN tail, containing 
the pulsar, with respect to the SNR shell yields 
$R_{\rm psr}/R_{\rm snr} < 0.677$. Therefore we conclude that the PWN inside 
SNR G327.1-1.1 is not bounded by a bow shock.

\section{Conlcusion}

We have performed a hydrodynamical simulation of a PWN, when the pulsar is
moving at a high velocity through its associated SNR. The simulation shows 
that the supersonic expandion stage of the PWN is terminated by the passage
of the reverse shock. Due to its high velocity, the pulsar is
positioned at the head of the PWN after the passage of the reverse shock.
The resulting morphology is similar to that of a PWN bow shock,
which is formed at half the crossing time when the pulsar is located at
a radius $R_{\rm psr}\simeq 0.677R_{\rm snr}$. We have applied these criteria
to the SNR G327.1-1.1 and concluded that the PWN inside this system has
not been deformed into a bow shock.


\end{document}